\pdfoutput=1
\documentclass[final,3p,times,twocolumn]{elsarticle}
\usepackage{blindtext, graphicx, amsmath, algorithm, algpseudocode, pifont, algcompatible, comment, layout, amsthm, amssymb}
\usepackage{enumitem}   
\usepackage{eso-pic}
\usepackage{booktabs}
\usepackage{float}
\usepackage{lscape}
\usepackage{subfig}
\usepackage[utf8]{inputenc}
\usepackage[english]{babel}
\usepackage{hyperref} 
\hypersetup{ colorlinks=true, linkcolor=black, filecolor=black, urlcolor=cyan, }

\usepackage{caption}
\journal{ICT Express}

\begin{document}

\begin{frontmatter}

\title{COVID-19 Detection through Deep Feature Extraction }
%\author{ Jash Dalvi\corref{cor1}, Aziz Bohra}
\author{\corref{cor1}Jash Dalvi}
\ead{jash.dalvi@somaiya.edu}

\author{Aziz Bohra}
\ead{aziz.bohra@somaiya.edu}

\address{Student, Department of Computer Engineering, K.J. Somaiya Institute of Engineering and Information Technology, Mumbai, India\\Student, Department of Computer Engineering, K.J. Somaiya Institute of Engineering and Information Technology, Mumbai, India}

\cortext[cor1]{Corresponding author}
%ead{jash.dalvi@somaiya.edu}

\begin{abstract}
The SARS-CoV2 virus, has caused a lot of tribulation to the human population. A predictive modeling, that can accurately determine whether a person is infected with COVID-19, is imperative. The study proposes a novel approach that utilizes deep feature extraction technique, pre-trained ResNet50 acting as the backbone of the network, combined with Logistic Regression as the head model. The proposed model has been trained on Kaggle COVID-19 Radiography Dataset. The proposed model achieves a cross-validation accuracy of 100\% on the COVID-19 and Normal X-Ray image classes. Similarly, when tested on combined three classes, the proposed model achieves 98.84\% accuracy. 
\end{abstract}

\begin{keyword}
SARS-CoV2 virus\sep  COVID-19 \sep ResNet50 
\end{keyword}

\end{frontmatter}

% \linenumbers

\section{Introduction}\label{sec1}

Coronavirus illnesses which are caused by the virus SARS-CoV2 have rapidly spread around the world. Although no particular theories or vaccinations exist for COVID-19, there are a number of clinical trials that are exploring a remedy to this problem. A lot of precautions have been suggested that can help in reducing the spread of the virus.

In this study, we provide a novel approach to detecting COVID-19 in X-Ray images. The model in the paper essentially consists of two stages:
\begin{enumerate}
  \item Feature Extraction through Pretrained ResNet50
  \item Classification through Logistic Regression
\end{enumerate}

Currently, a lot of deep learning models thrive on copious amounts of data. But due to the paucity of X-Ray image data related to COVID-19 we have suggested a novel approach in this paper. The model shown in the paper helps to overcome the problem of relatively low quantity of data.

The paper is organized as follows. Section 2 presents a Literature Survey of the models developed till now to tackle the same problem. Section 3 presents the methodology associated with the model. Section 4 presents results and findings. Section 5 concludes the paper.

\begin{figure*}[h]
\centering
\captionsetup{justification=centering,margin=2cm}
{{\includegraphics[width=\linewidth]{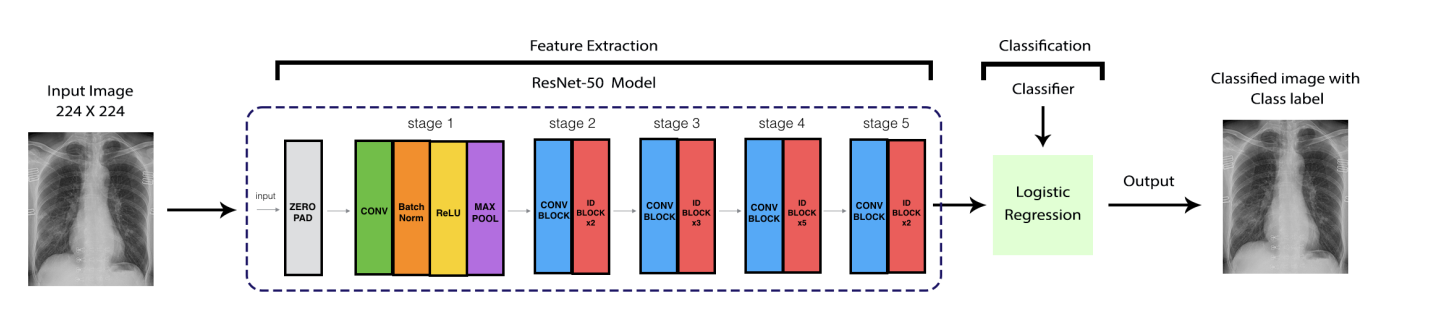} }}
\caption[justification=center]{Proposed Architecture-ResNet50 + Logistic Regression }
\end{figure*}

\section{Literature Survey}\label{sec2}
Various deep learning architectures have been employed as a tool to automate the detection of COVID-19 \cite{polsinelli2020light,li2020robust}. Covid-Net from \cite{wang2020covid} is trained on the COVIDx dataset and yields an accuracy of 93.30\% with an F1 score of 98.90\% and a Sensitivity of 91.00\%. Researchers in \cite{narin2021automatic} developed a Deep Learning model using InceptionV3, ResNet50, ResNet101, ResNet152, Inception-ResNetV2 which has an accuracy (mean) of 98\%. This model is trained on the ChestX-ray8 dataset which constituted 96\% Sensitivity, 98\% F1 score, 100\% Specificity, and 100\% Precision. \cite{taresh2020transfer} has incorporated CNN with Transfer Learning where two models have been trained namely VGG19 and Mobile-Net. Mobile Net outperformed VGG19 and proved to be one of the best models in the detection of COVID-19 from X-Ray images with an accuracy of 98.28\%. It was trained on a dataset that was gathered from various sources. The above two models from \cite{taresh2020transfer} were compared with other CNN models like Inception, Xception, and Inception ResNet v2, and yet they performed better with Sensitivity around 98.66 and a Specificity of 96.46. Authors of \cite{sethy2020detection} have used the GitHub repository of Dr. Joseph Cohen and Kaggle X-Ray images of pneumonia. \cite{sethy2020detection} uses CNN models with SVM at the end. The results revealed that ResNet50 + SVM proved to be the best model with an accuracy of 95.33\% and the same Sensitivity, 95.34\% was the Specificity. Researchers of \cite{hemdan2020covidx} have concluded that VGG19 and Dense-Net showed good results and they were evaluated on the 80-20\% training and validation proportions. Results illustrated an accuracy of 90\% and an F1 score of 91\% along with 100\% Precision and Sensitivity. The Confidence-Aware Anomaly Detection (CAAD) model \cite{zhang2020viral} bags an accuracy of 96\%, trained on X-Viral and X-Covid datasets (X-Ray images). Covid-ResNet from \cite{farooq2020covid} is developed using pre-trained weights from ImageNet incorporating them with the 50 layers of the Residual Network (ResNet). This model is also trained on the COVIDx dataset and has an accuracy of 96.23\% along with a 100\% F1 score and Precision.

\section{Methods and Data}\label{sec2}

\subsection{Data Description}
The proposed study has used the COVID-19 Radiography database \cite{rahman2021exploring,chowdhury2020can} which is available on Kaggle. A team of researchers from Qatar University, Doha, and the University of Dhaka, Bangladesh along with their collaborators from Pakistan and Malaysia in collaboration with medical doctors have created a database of chest X-ray images for COVID-19 positive cases along with Normal and Viral Pneumonia images. The dataset includes approximately 1200 images of each class\ - COVID-19, Normal and Viral Pneumonia. Researchers can find this dataset at the following link: \url{https://www.kaggle.com/tawsifurrahman/covid19-radiography-database}

\subsection{ResNet Description}

To increase the depth of the Neural Network (NN) \cite{wang2003artificial}, directly stacking layers one after the other will not work out. The deeper the network, the more difficult it gets to train it. This difficulty is caused by the diminishing gradient - as it is propagated backwards to the prior layers, multiplicative repetition leads to extremely small values for the gradient. Consequently, performance gets impeded on account of deeper NN.

Skip Connection, a concept introduced by the Residual Networks mainly known as ResNets \cite{he2016deep}. In back-propagation, this concept provides an alternative path for the gradient. Experiments have evaluated that such additional paths benefit the model to converge. Skipping layers in the NN and feeding the output as an input to the further layers (excluding the immediate next layer) assists the whole NN with performance and obviously minimizes the situation of vanishing gradients.
\begin{equation}
 \frac{\partial L}{\partial x} =  \frac{\partial L}{\partial H}  \cdot\frac{\partial H}{\partial x} 
 = \frac{\partial L}{\partial H} \left( \frac{\partial F}{\partial x} + 1 \right) 
 =  \frac{\partial L}{\partial H}  \cdot\frac{\partial F}{\partial x} + \frac{\partial L}{\partial H} 
\end{equation}

\begin{figure}
\centering
{{\includegraphics[width=6cm]{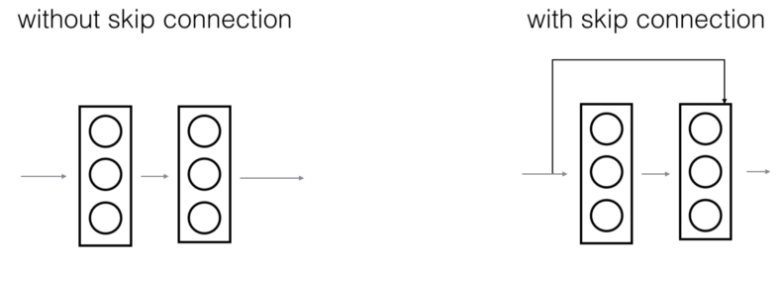} }}%
\caption{Skip Connection}
\end{figure}

One of the variants of the ResNet is ResNet-50 which constitutes single Max-Pool and Average-Pool layers along with 48 Convolution layers. In addition, it has \(3.8 x 10^9\) Floating point operations.

\subsection{Logistic Regression}
A logistic regression model processes a weighted sum of the independent variables (input features) along with the bias term and outputs the log of the result, unlike linear regression.
The logistic (sigma) is a sigmoid function (S-shaped) that outputs a number between 0 and 1. 

When the target (dependent variable) is categorical, logistic regression proves beneficial. If the output probability is greater than 0.5 then the model predicts that the instance belongs to the particular class and it is considered as 1, otherwise 0. Thus, it is a binary classifier.

\begin{figure}[h]
\centering
\subfloat[\centering 
]{{\includegraphics[width=7cm]{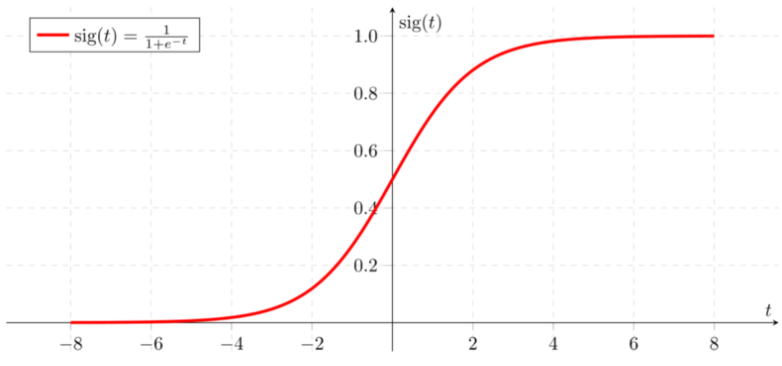} }}
\caption{Sigmoid function in Logistic Regression}
\end{figure}

\subsection{Methodology}
In this paper, the feature extraction technique is proposed. Feature extraction with respect to deep learning consists of using a CNN as a Base model and Machine learning model as a head model. The Convolutional Neural Network \cite{gu2018recent} used in this case is ResNet-50, pretrained on the ImageNet Dataset. The Fully Connected Layers of the ResNet-50 Model are removed. The whole dataset is passed through ResNet-50, which has its Fully Connected layers removed. Each image gets transformed into a vector of shape, 7 * 7 * 2048. We flatten this output vector into a one-dimensional vector of shape 100352, and save the combined vector of the whole dataset. The second part comprises using these vectors as input to the Machine learning model. The model used in this case is logistic regression. As Vanilla Neural Networks thrive on larger data, there is a limitation with respect to generalization. The proposed model utilizes a hybrid approach, CNN + ML base model, which helps in generalization.

\section{Results}
The dataset on which the model was tested consists of three classes, Normal, COVID, and Pneumonia. The Normal class has 1341 images, the COVID class has 1200 images, and the Pneumonia class has 1345 images. The Resnet50 + logistic regression model with different hyperparameters and considering different classes was compared. The hyperparameter ‘C’ was considered, which stands for the inverse of regularization strength. Regularization generally refers to the concept that there should be a complexity penalty for more extreme parameters. The idea is that just looking at the training data and not paying attention to how extreme one's parameters are leads to overfitting. A high value of ‘C’ tells the model to give high weight to the training data, and a lower weight to the complexity penalty. A low value tells the model to give more weight to this complexity penalty at the expense of fitting to the training data.

\begin{figure}[h]
\centering
\subfloat[\centering Confusion matrix with two classes
  %\begin{minipage}[b]{0.4\textwidth}
]{{\includegraphics[width=3.3cm]{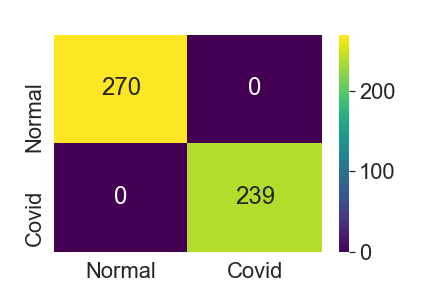} }}%
    %\caption{Flower one.}
  %\end{minipage}
  \hfill
  \subfloat[\centering Confusion matrix with three classes
  %\begin{minipage}[b]{0.4\textwidth}
]{{\includegraphics[width=3.3cm]{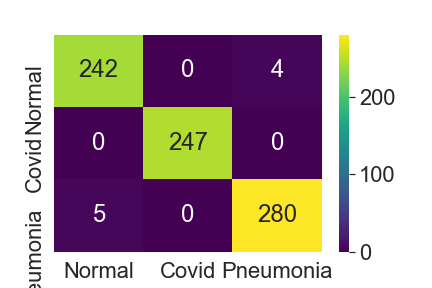} }}%
    \caption{Confusion matrices of ReNet50 + LR model with C = 0.1}
  %\end{minipage}
\end{figure}

% \begin{figure}[h]
% \centering
% \subfloat[\centering label 1]{{\includegraphics[width=3.3cm]{LogisticRegression1_cm.png} }}%
%     \qquad
%     \subfloat[\centering label 2]{{\includegraphics[width=3.3cm]{LogisticRegression1_covid_cm.png} }}%
%     \caption{2 Figures side by side}%
%     \label{fig:ex}
% \end{figure}

\iffalse

\includegraphics[width=.4\columnwidth]{LogisticRegression1_cm.png}
\caption{Example of a figure.}
\label{fig:ex}
\end{figure}

\begin{figure}
\centering
\includegraphics[width=.4\columnwidth]{LogisticRegression1_covid_cm.png}
\caption{Example of a figure.}
\label{fig:ex}
\end{figure}
\fi

\begin{table*}
\centering
\captionsetup{justification = centering}
\caption{Results of different studies}
\label{table:1}
\begin{tabular}{||c c c c c c||} 
 \hline
\toprule
Paper & Images & Accuracy & Sensitivity & F1 Score & Precision\\ 
\midrule
\cite{wang2020covid}               &258       &93.00 &91.00 &98.00 &-\\
\cite{narin2021automatic} &100         & 98.00        & 96.00     &98.00 &100.00  \\
\cite{taresh2020transfer}  & 3886       &98.28 &98.66 &- &-        \\
\cite{sethy2020detection}  & 381        & 95.33 &95.33 &95.34 & -        \\ 
\cite{hemdan2020covidx}  & 50 &90.00 &100.00 &91.00 &100.00 \\
\cite{zhang2020viral} & 1531 & 96.00 & - & - & - \\
\cite{farooq2020covid}  & 13975 & 96.23 & - & 100.00 & - \\
Proposed study & 3886 & 98.84 & 99.00 & 99.00 & 99.00 \\
\bottomrule
\end{tabular}
\end{table*}

The different values of ‘C’ considered were 0.001,0.01,0.1,1. The same model was trained and tested considering different classes. In one case, all the three classes were considered, Normal, COVID, and Pneumonia, while in the other case only the Normal and COVID class was taken into consideration. The data was split into training and testing data. The training data comprised 80\% of the total data, while the testing data comprised 20\% of the whole data. The dataset was split equally into three classes in order to avoid the bias that can generate from unequal splitting.

The best accuracy was obtained with ResNet50 + logistic regression with hyperparameter C equal to 0.1. Different evaluation metrics are used to analyze the COVID-19 model’s performance. Among them, the most used metrics for the detection of COVID-19 are Accuracy, Sensitivity, F1 score, Specificity, and Precision.  The performance of the proposed method in this study is evaluated with these metrics. The test accuracy when three classes, Normal, COVID, and Pneumonia, were considered was 98.84\%. Similarly, the test accuracy when two classes, Normal and COVID, were considered was 100\%. For different values of hyperparameter C, 0.001 ,0.01,1, the testing accuracy  considering three classes was 98.45\%, 98.71\%, and 98.84\% respectively. For similar values, the testing accuracy considering three classes was 98.80\%, 100\%, and 100\% respectively. From the accuracy values, it is evident that the proposed model performs efficiently as well as generalizes better than any other previous models.

\section{Conclusion and Discussion}
In this study, model detecting COVID-19 constitutes a deep learning model along with Logistic Regression. The model was trained on Kaggle’s Radiography dataset where the X-Ray images were put through the ResNet-50 deep learning model. Grid Search was adopted for fine tuning of hyper-parameters and cross-validation was performed on the whole dataset. The study mentioned above gives much better results compared to other related works in the detection of COVID-19. Our model accounts for an accuracy of 98.84\% for all three classes (Normal, COVID-19, Pneumonia) and has 99.00\% of Precision, Recall, and an F1 score. All the other models observed an accuracy lower than the proposed model in this study.

Due to the limitation of the size of the dataset, the model is developed using approximately 1200 X-Ray images per class. This model can be utilized and tested on a subset of the population before generalizing on a larger part of the population. The performance and reliability of this model can be enhanced using a larger dataset.

In conclusion, it is evident that such techniques can be very useful in the detection of COVID-19 and Pneumonia. The test results prove that the proposed model may be a helping aid to the whole medical staff all-round the globe and also paves way for other such medical aids.

\section*{Conflict of interest}
The authors declare that there is no conflict of interest in this paper.

%% References


\bibliographystyle{elsarticle-num}

\vspace{-0.3cm}

\bibliography{sample}
\end{document}